# Orbit Maneuver of Spinning Tether via Tidal Force*

Hexi Baoyin†, Yang Yu and Junfeng Li
*School of Aerospace, Tsinghua University, Beijing, 100084, China*

**Recently, the spinning tethered system is regarded as a typical and fundamental space structure attracting great interest of the aerospace engineers, and has been discussed primarily for specific space missions in past decades, including on-orbit capture and propellantless orbit transfer etc. The present work studies the dynamical behaviours of a fast spinning tethered binary system under central gravitational field, and derives principles of the basic laws of orbital maneuver. Considering the characteristics of coupled librational and orbital motions, an averaging method is introduced to deal with the slow-fast system equation, thus a definite equivalent model is derived. The general orbit motion is completely determined analytically, including the orbit geometry, periodicity, conversations and moving region etc. Since the possibility of orbit control using tether reaction has been proved by previous studies, special attention is paid to the transportation mode of angular momentum and mechanical energy between the orbit and libration. The effect of tether length change on the orbit shape is verified both in the averaged model and original model. The results show the orbit angular momentum and mechanical energy can be controlled independently, and the operating principles of tether reactions are derived for special modification of orbit shape.**

## Nomenclature

| | | |
|---|---|---|
| $R,\boldsymbol{R}$ | = | radius and radius vector with respect to the center of gravity |
| $l,\boldsymbol{l}$ | = | tether length and the vector between the binary satellites |
| $F, \boldsymbol{F}$ | = | gravitational force |
| $\boldsymbol{T}$ | = | tensile force on the binary satellite |
| $\lambda$ | = | normalized mass of binary satellite |
| $\mu$ | = | gravitational coefficient |
| $\rho$ | = | polar radius of mass center |
| $\varphi$ | = | polar angular of mass center |

* This work was supported by the National Natural Science Foundation of China (No. 11072122)
† Corresponding author, Email: baoyin@tsinghua.edu.cn

$\alpha$ = azimuth of tether
$M$ = angular momentum
$E$ = mechanical energy
$\Delta$ = difference between the follower and the leader
$a$ = semi-major axis
$e$ = eccentricity ratio
$\omega$ = argument of perigee
$f$ = true anomaly
$n$ = mean rate
$p$ = semi-latus rectum

*Subscripts*

$C$ = system mass center
$i$ = 1, 2
$avg$ = averaged magnitude
$r$ = radial component

## I. Introduction

THERE are rich literature researching the coupling effects between librational and orbital motions of spacecraft, that the tidal force on large-scale spacecraft might play a considerable role in the attitude dynamics and control[1,2]. Since tethered satellites seem the only artificial spacecraft with up to ten thousands of kilometers' dimensions, this coupling connection between their orbits and attitudes is relatively strong in the space motion. In addition, the inertia momentum of tethered system is usually controllable because the tether length is under control of the hoisted machine fixed on the satellite. Therefore a converse idea was proposed that orbit transfer might be realized for tethered system just by tether length reaction[3,4]. In 1987, Martinez-Sanchez et al. of MIT formulated a scheme of orbital modifications using tether-length variation once per orbit. Additionally, the tether-length control laws are developed and discussed for several specific applications such as orbit modification of the space station[5]. Landis of NASA studied the propellantless way of propelling tethered system using tether deployment and retrievement, in which the method of reaction against the gravitational gradient was advocated and the first order series were adopted to estimate the efficiency of orbital propulsion[6]. Gratus et al. of Lancaster University gave a formula for the rate of changing the tether-length using a series of reasonable approximations, and their formula predicts that by suitably varying the length of a 50 km connecting tether, the system can rise at about 300 meters an hour in low earth orbit[7].

Unlike propellantless propulsion in other senses such as solar sails, magnet sails, electronic sails and electro-dynamical tethers, this kind of orbital propulsion of tethered system depends on no physical mechanisms other than the Newtonian gravitational field[8,9]. The curiosity is that the desired orbit changes could be achieved only by deploying and retrieving the tether constantly, seeming like external movement being dominated by interactions inside the system.

In this paper, this intriguing phenomenon is investigated in detail with a spinning tethered binary system model. First, the dynamical behaviours of spinning tethered binary system are analyzed for the delicate connection between the librational and orbital motions. With the fast variant filtered by averaging approach, the general orbital motion of mass center is completely determined with an analytical method employed. The effects of tether reaction (including deployment and retrievement) are discussed for temporary and secular impact on the orbital motion, in which the mechanism of energy exchange and angular momentum transportation are concentrated. Then, a perturbation method is applied to investigate the relationship between the orbit shape variation and the tether length change. Some basic laws of motion control in gravitational field are obtained with the observation of the perturbation equations and verified by numerical simulations. Finally, calculation results show several special applications of this propulsion method in orbit maneuvres.

## II. Equations, Conversation and stability of Motion

As shown in Fig.1, a spinning binary system consists of two mass points connected by a massless tether, and is modelled as a simple and typical structure reserving the essential characteristics of the system. The governing equations of orbital and librational motion are obtained in this section and the conversation quantities are represented in geometric way.

Considering the tethered binary system simplified as two mass particles connected via a lightweight tether, the motion of this system could be resolved into orbital motion and librational motion around the mass center. Since the attitude motion out of orbit plane is decoupling with that in the plane, all the motion discussed in this paper is delimited in the orbit plane.

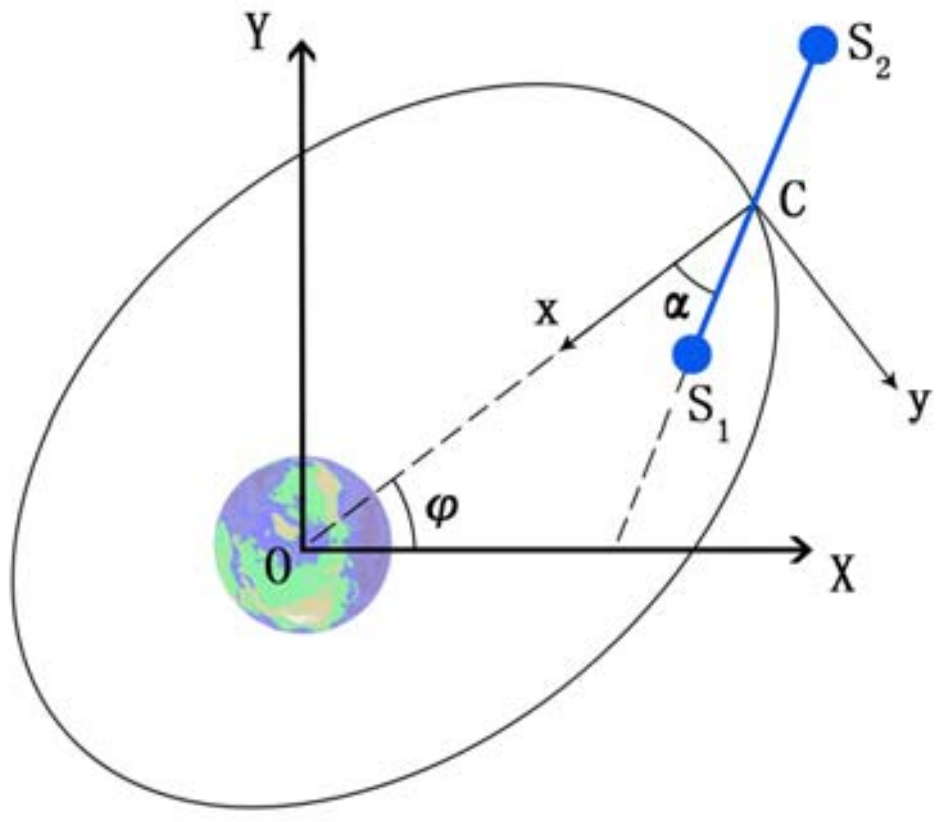

**Fig. 1 The spinning tethered binary system model. The blue object is the tethered system; solid arrows indicate the reference systems and dashed lines are guides for illustration.**

As Fig. 1 illustrates, $S_i$ are the mass points. Coordinates OXY is an inertial frame in the orbit plane, and Cxy is the orbit coordinate frame, in which C is the mass center, axis Cx points to the gravitation center and axis Cy points toward the flying direction, forming right-handed system.

For $S_i$, it yields

$$\lambda_i \ddot{\boldsymbol{R}}_i = \boldsymbol{F}_i + \boldsymbol{T}_i \tag{1}$$

Combining the formulas $\boldsymbol{R}_C = \sum \lambda_i \boldsymbol{R}_i$ , $\sum \boldsymbol{T}_i = \boldsymbol{0}$ , $\boldsymbol{F}_i = -\mu\lambda_i \boldsymbol{R}_i / R_i^3$ and $\boldsymbol{l} = \boldsymbol{R}_2 - \boldsymbol{R}_1$ , the vector equations of motion are deduced.

$$\ddot{\boldsymbol{R}}_C = -\sum \mu\lambda_i \frac{\boldsymbol{R}_i}{R_i^3} \tag{2}$$

$$\boldsymbol{l} \times \ddot{\boldsymbol{l}} = \boldsymbol{l} \times \left( -\mu \frac{\boldsymbol{R}_2}{R_2^3} + \mu \frac{\boldsymbol{R}_1}{R_1^3} \right) \tag{3}$$

And it can be written in component form,

$$\ddot{\rho} - \rho\dot{\varphi}^2 = -\mu\lambda_1 \frac{\rho - \lambda_2 l \cos\alpha}{\left[\rho^2 - 2\rho\lambda_2 l \cos\alpha + (\lambda_2 l)^2\right]^{3/2}} - \mu\lambda_2 \frac{\rho + \lambda_1 l \cos\alpha}{\left[\rho^2 + 2\rho\lambda_1 l \cos\alpha + (\lambda_1 l)^2\right]^{3/2}} \tag{4}$$

$$\rho\ddot{\varphi} + 2\dot{\rho}\dot{\varphi} = \frac{\mu\lambda_1\lambda_2 l \sin\alpha}{\left[\rho^2 - 2\rho\lambda_2 l \cos\alpha + (\lambda_2 l)^2\right]^{3/2}} - \frac{\mu\lambda_1\lambda_2 l \sin\alpha}{\left[\rho^2 + 2\rho\lambda_1 l \cos\alpha + (\lambda_1 l)^2\right]^{3/2}} \tag{5}$$

$$l(\ddot{\varphi} + \ddot{\alpha}) + 2\dot{l}(\dot{\varphi} + \dot{\alpha}) = \frac{-\mu\rho \sin\alpha}{\left[\rho^2 - 2\rho\lambda_2 l \cos\alpha + (\lambda_2 l)^2\right]^{3/2}} + \frac{\mu\rho \sin\alpha}{\left[\rho^2 + 2\rho\lambda_1 l \cos\alpha + (\lambda_1 l)^2\right]^{3/2}} \tag{6}$$

Considering the system angular momentum is always conserved, and the mechanical energy is conserved only with invariant tether length, the conservation quantities can be obtained as

$$\rho^2\dot{\varphi} + \lambda_1\lambda_2 l^2 (\dot{\varphi} + \dot{\alpha}) = M \tag{7}$$

$$\dot{\rho}^2 + (\rho\dot{\varphi})^2 + \lambda_1\lambda_2 l^2 (\dot{\varphi} + \dot{\alpha})^2 = \frac{2\mu\lambda_1}{\sqrt{\rho^2 - 2\rho\lambda_2 l \cos\alpha + (\lambda_2 l)^2}} + \frac{2\mu\lambda_2}{\sqrt{\rho^2 + 2\rho\lambda_1 l \cos\alpha + (\lambda_1 l)^2}} + 2E \tag{8}$$

The constants $M$ and $E$ in Eqs.(7), (8) respectively represent the total angular momentum and mechanical energy of the system. Right hand side of Eq.(8) could be approximated by $2\mu/\rho + 2E$ . In velocity subspace $\{\rho\dot{\varphi}, l(\dot{\varphi}+\dot{\alpha}), \dot{\rho}\}$, formula (7) and (8) describe two simple geometric objects that are a plane and an ellipsoid respectively, with their slope and dimension varying constantly. And the intersection ellipse displays the feasible set for real motion, as shown in Fig. 2.

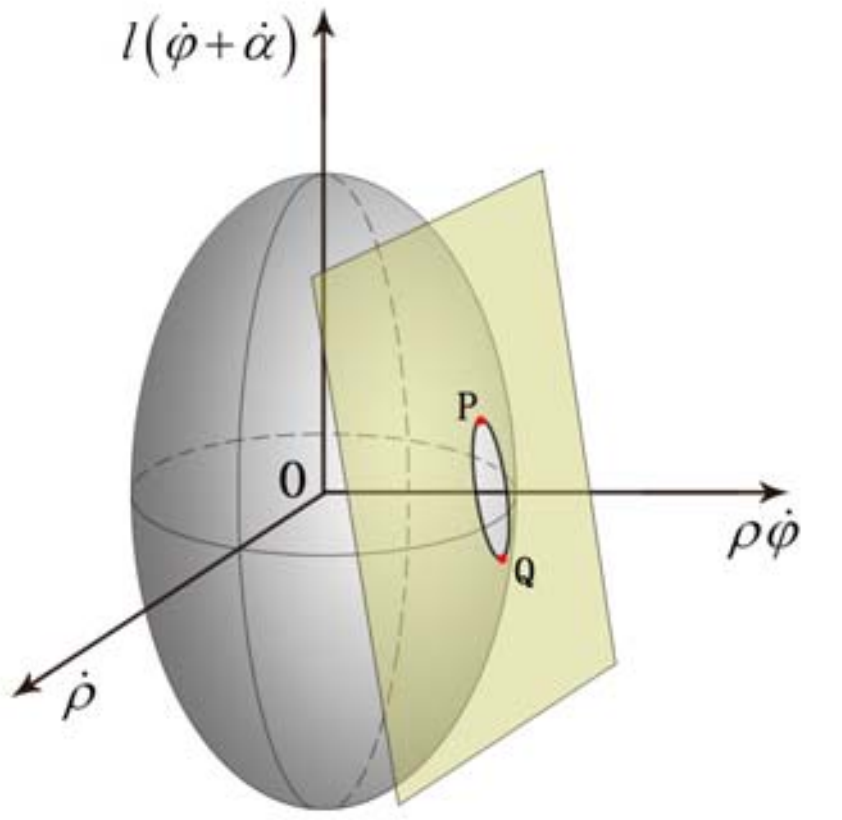

**Fig. 2 Motion in velocity subspace. The right ellipse presents the feasible set of real motion, and for a fast spinning tether i.e. $|\dot{\alpha}| \gg |\dot{\varphi}|$, the feasible values concentrated around the vertexes P (for the cases that spinning and orbital directions are consistent) and Q (for the cases that spinning and orbital directions are opposite).**

And for a fast spinning tether, i.e. $|\dot{\alpha}| \gg |\dot{\varphi}|$, the feasible values concentrated around the vertexes of this ellipse, where the radial velocity $\dot{\rho}$ equals zero. It is verified these vertexes are central equilibrium points in phase space, which indicate the stability of the motion without tether reaction.

## III. Long-term Orbital Motion

In this section, the long-term orbital motion of the system mass center is investigated with an analytic method presented in [10]. Considering the spinning angular rate is much larger than the orbital angular rate, the long-periodic motion and short-periodic motion can be treated separately. The spin angle $\alpha$ is fast changing and weakly coupled with the orbital motion, thus an averaging method is used to isolate the long-term orbital motion of system mass center. Based on the slow-fast decomposition theory of singular perturbation, the reduced subsystem approximates to original system.

The integrals of equations (4), (5) for spinning period respect to $\alpha$ are represented as

$$\ddot{\rho}-\rho\dot{\varphi}^2=-\frac{\mu}{\pi\rho^2}\left[\lambda_1 F(\eta_2)+\lambda_2 F(-\eta_1)\right] \tag{9}$$

$$\rho\ddot{\varphi}+2\dot{\rho}\dot{\varphi}=0 \tag{10}$$

Where $\eta_i=\frac{\lambda_i l}{\rho}$ , $F(x)=\frac{1}{1+x}EllipticK\left(\frac{2\sqrt{x}}{1+x}\right)+\frac{1}{1-x}EllipticE\left(\frac{2\sqrt{x}}{1+x}\right)$ and $EllipticK$ , $EllipticE$ are complete elliptic integrals of first and second kind respectively. Denoting right hand side of equation (9) as $F_C(\rho)$, it determines a central field which is deferent from that derived by a mass particle. The system equations (9), (10) are conserved without considering tether reaction. Defining $Z(x)=EllipticK\left(2\sqrt{x/(1+x)^2}\right)/(1+x)$ and $U=-\frac{2\mu}{\pi\rho}\left[\lambda_1 Z(\eta_2)+\lambda_2 Z(-\eta_1)\right]$, equation (9) can be rewritten as

$$\ddot{\rho}-\rho\dot{\varphi}^2=-\frac{\partial U}{\partial\rho} \tag{11}$$

Combining the conservation of angular momentum and mechanical energy, the subsystem (10) and (11) could be further reduced with the method presented in [8].

$$\ddot{\rho}=-\frac{\partial V}{\partial\rho} \tag{12}$$

where $V(\rho)=U(\rho)+\frac{M^2}{2\rho^2}$, and the conservation equation can be obtained

$$\frac{1}{2}\dot{\rho}^2+V(\rho)=E \tag{13}$$

Fig. 3 illustrated the variation of effective potential energy $V(\rho)$ with a certain initial condition. The range of polar radius $\rho$ is limited by $V(\rho)\le E$, based on the simple fact that $\dot{\rho}^2/2=E-V(\rho)$ is always positive, thus the orbits of mass center lie in the annular region $0<R_{\min}\le\rho\le R_{\max}<\infty$ with $R_{\min}$, $R_{\max}$ determined by $V(\rho)=E$.

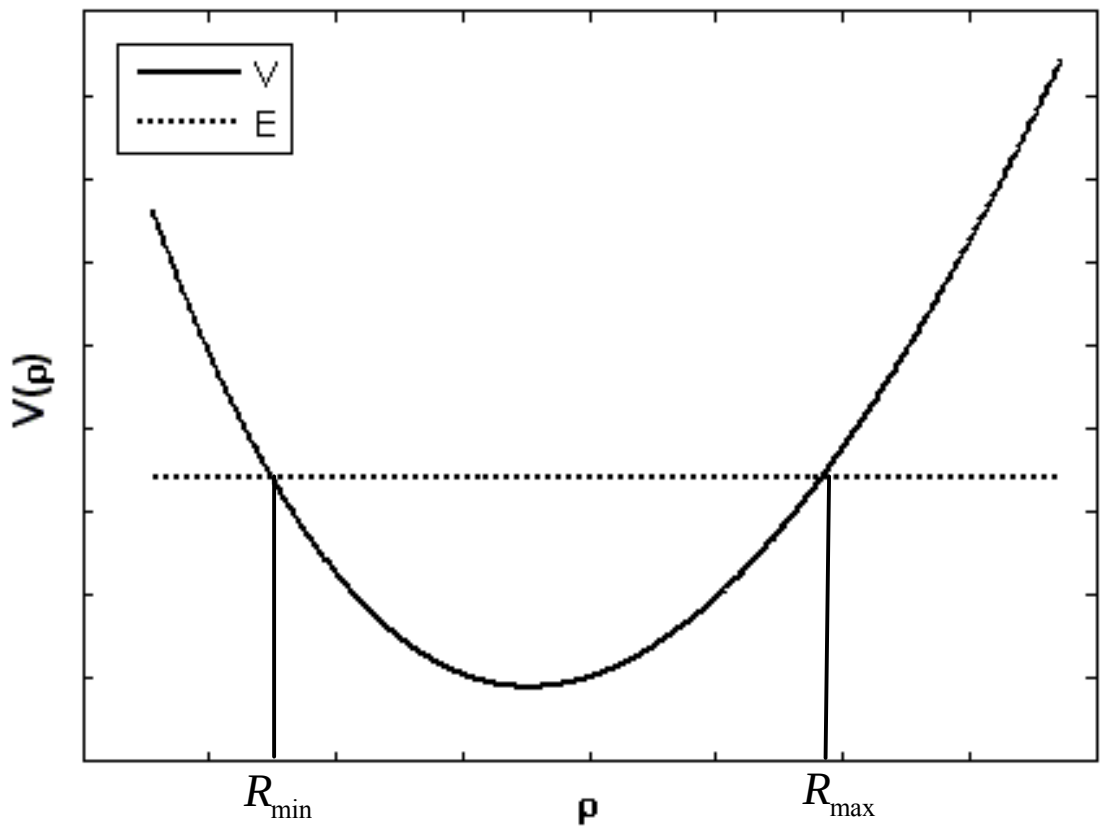

**Fig. 3 The effective potential energy V(ρ) varying with ρ**

Since the boundaries of the orbits are determined, the pattern of orbital motion could be further explored. The polar radius $\rho$ oscillates periodically between $R_{\min}$ and $R_{\max}$, and the polar angle $\varphi$ varies monotonically as time, which is illustrated in Fig. 4.

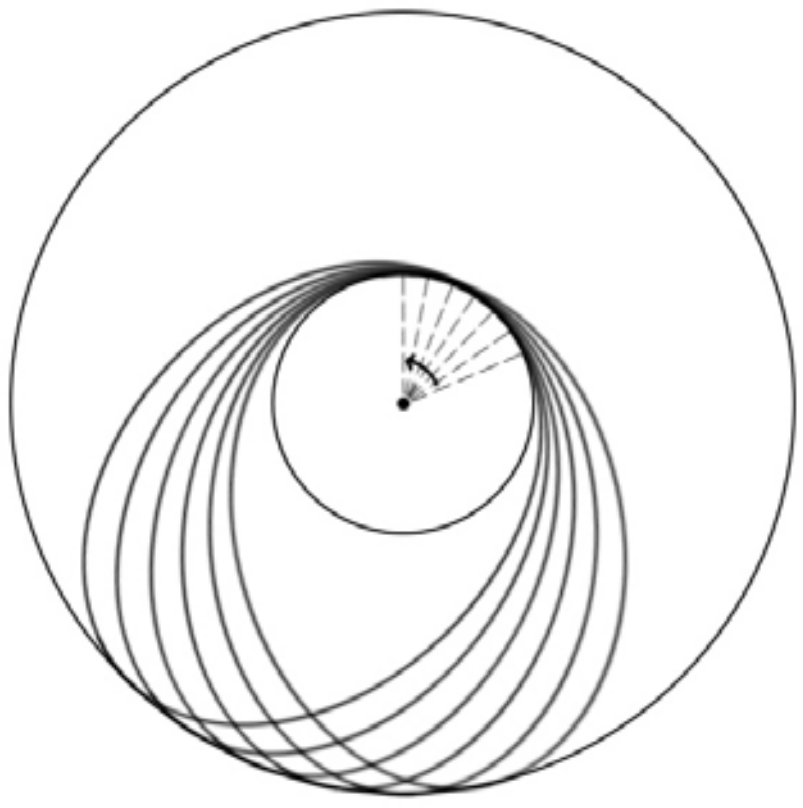

**Fig. 4 The orbital motion over long term**

The orbit is symmetrical about each ray leading from center to perigee or apogee because of the symmetry of motion equations about the variant $\varphi$. Concentrating on the long-term evolution of orbit, the drifting of perigee could be estimated analytically according to [10]. As Fig.4 illustrated, this drifting is unidirectional and the drifting rate is homogenous. Based on formula (14), (15), the orbit drifts clockwise if $\Phi > \pi$ and anti-clockwise if $\Phi < \pi$; the drifting rate of perigee was estimated by

$$\Phi = \int_{R_{\min}}^{R_{\max}} \frac{M d\rho}{\rho^2 \sqrt{2\left(E - V(\rho)\right)}} \tag{14}$$

$$\nu = \left[ \pi - \int_{R_{\min}}^{R_{\max}} \frac{M d\rho}{\rho^2 \sqrt{2\left(E - V(\rho)\right)}} \right] \Bigg/ \int_{R_{\min}}^{R_{\max}} \frac{d\rho}{\sqrt{2\left(E - V(\rho)\right)}} \tag{15}$$

Where the denominator is the period from perigee to apogee, and as calculated, the drifting rate of perigee increases evidently with the values of $M$ and $E$. Additionally, the closure of orbit also depends on $\Phi$ : the orbit is closed if the angle $\Phi$ is commensurable with π, i.e., if $\Phi = 2\pi(m/n)$ where *m* and *n* are integers.

## IV. Short-term Effects of Tether Reaction

This section take a detailed look at the mechanism of angular momentum transfer and mechanical energy input/output, for which the local effects of tether reaction are inspected in short term. And these effects could be advantageous for the long-term motion control with tether deploying and retrieving operations.

### A. Local Energy Effects

During a monotonous tether length reaction, which is deployment or retrievement, the tensile force in tether produces an abrupt change of the system mechanical energy. For a fast spinning tether, this energy change depends on the tether length variation. According to equation (1), the tether vector satisfies

$$\ddot{\boldsymbol{l}} = -\mu \frac{\boldsymbol{R}_2}{R_2^3} + \mu \frac{\boldsymbol{R}_1}{R_1^3} + \frac{\boldsymbol{T}_2}{m_2} - \frac{\boldsymbol{T}_1}{m_1} \tag{16}$$

In component form, equation (16) is represented as

$$\ddot{l}-l(\dot{\varphi}+\dot{\alpha})^2=\frac{\mu(\rho\cos\alpha-\lambda_2 l)}{\left[\rho^2-2\rho\lambda_2 l\cos\alpha+(\lambda_2 l)^2\right]^{3/2}}-\frac{\mu(\rho\cos\alpha+\lambda_1 l)}{\left[\rho^2+2\rho\lambda_1 l\cos\alpha+(\lambda_1 l)^2\right]^{3/2}}-\frac{T}{\lambda_1\lambda_2 M} \quad (17)$$

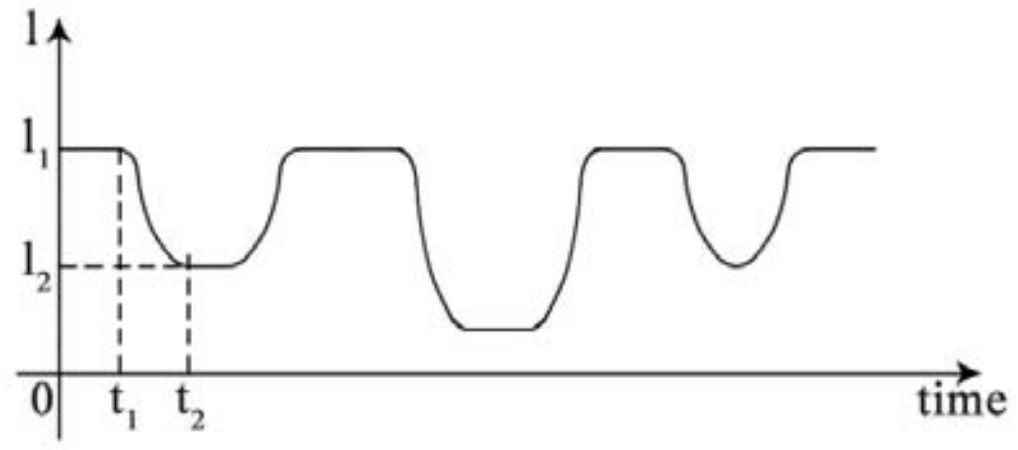

**Fig. 5 The variation of tether length as time**

Fig. 5 illustrated the variation of tether length as time, with deployment and retrievement alternated. Between the deployment and retrievement, the tether length keeps invariant. As $t_i$, $l_i$ defined in the figure, the mechanical energy change $\Delta E=-\int_{t_1}^{t_2}T\cdot\dot{l}dt=-\int_{l_1}^{l_2}Tdl$. Combining with equation (17), it makes

$$\int_{t_1}^{t_2}\ddot{l}\cdot\dot{l}dt-\int_{l_1}^{l_2}l(\dot{\varphi}+\dot{\alpha})^2dl=\int_{l_1}^{l_2}\left\{\frac{\mu(\rho\cos\alpha-\lambda_2 l)}{\left[\rho^2-2\rho\lambda_2 l\cos\alpha+(\lambda_2 l)^2\right]^{3/2}}-\frac{\mu(\rho\cos\alpha+\lambda_1 l)}{\left[\rho^2+2\rho\lambda_1 l\cos\alpha+(\lambda_1 l)^2\right]^{3/2}}\right\}dl+\frac{\Delta E}{\lambda_1\lambda_2 M} \quad (18)$$

Where the first term of LHS $\int_{t_1}^{t_2}\ddot{l}\cdot\dot{l}dt=\frac{1}{2}\dot{l}^2\Big|_{t_1}^{t_2}=0$ , and the second term

$$\int_{l_1}^{l_2}l(\dot{\varphi}+\dot{\alpha})^2dl=\int_{l_1}^{l_2}\left(\frac{\rho^2\dot{\varphi}-M_0}{\lambda_1\lambda_2}\right)^2\frac{1}{l^3}dl \quad (19)$$

Noticing the orbital angular momentum $\rho^2\dot{\varphi}$ is much larger than spinning angular momentum $\lambda_1\lambda_2 l^2(\dot{\varphi}+\dot{\alpha})$ and varies much slowly, the formula (19) indicates the following facts: tether deployment makes an abrupt energy decrease; on the contrary, tether retrievement makes an abrupt energy increase; the tether reaction at higher attitude of the orbit leads to larger energy change; more energy increase can be achieved with the same tether length variation only by improving the spinning angular momentum.

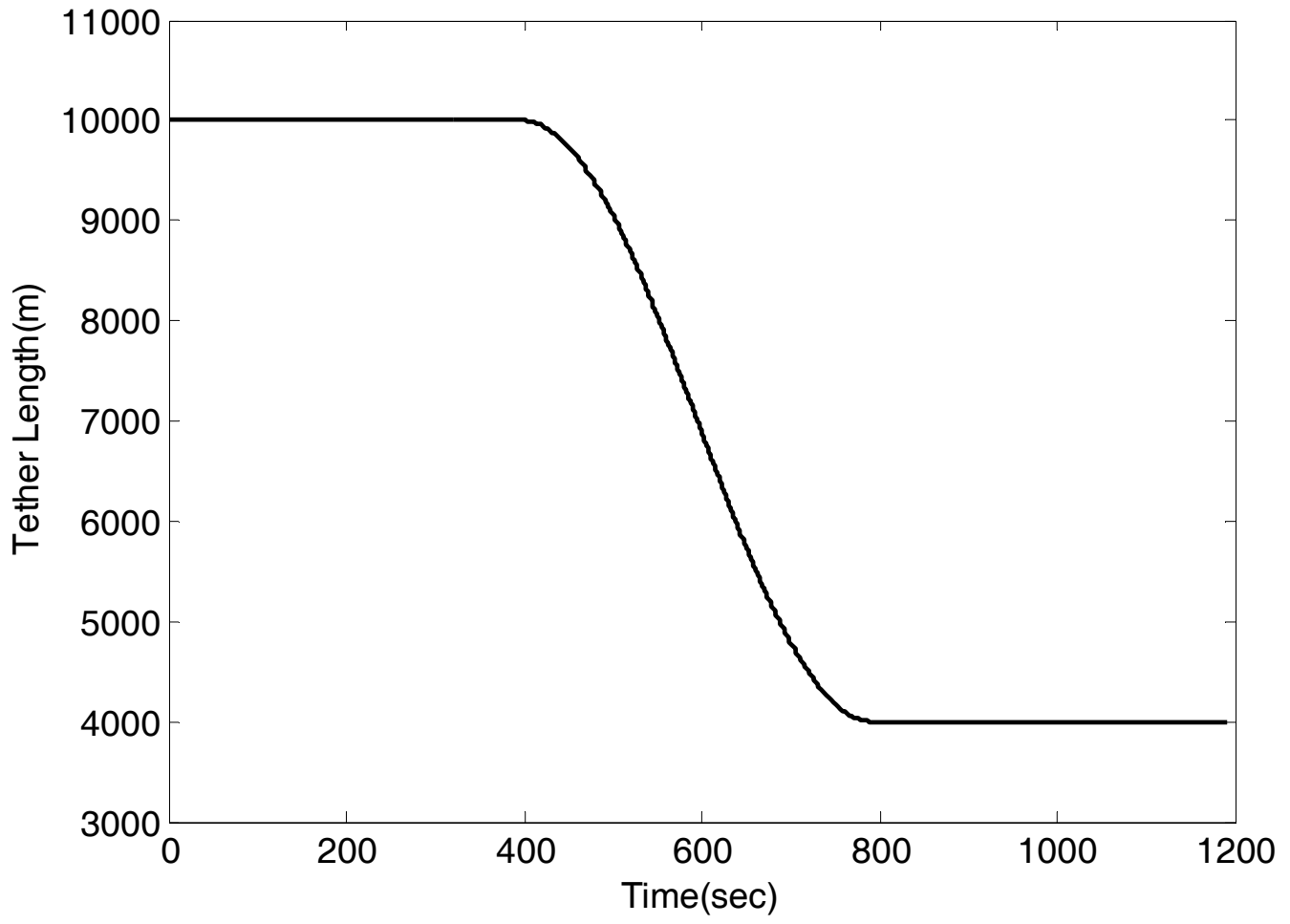

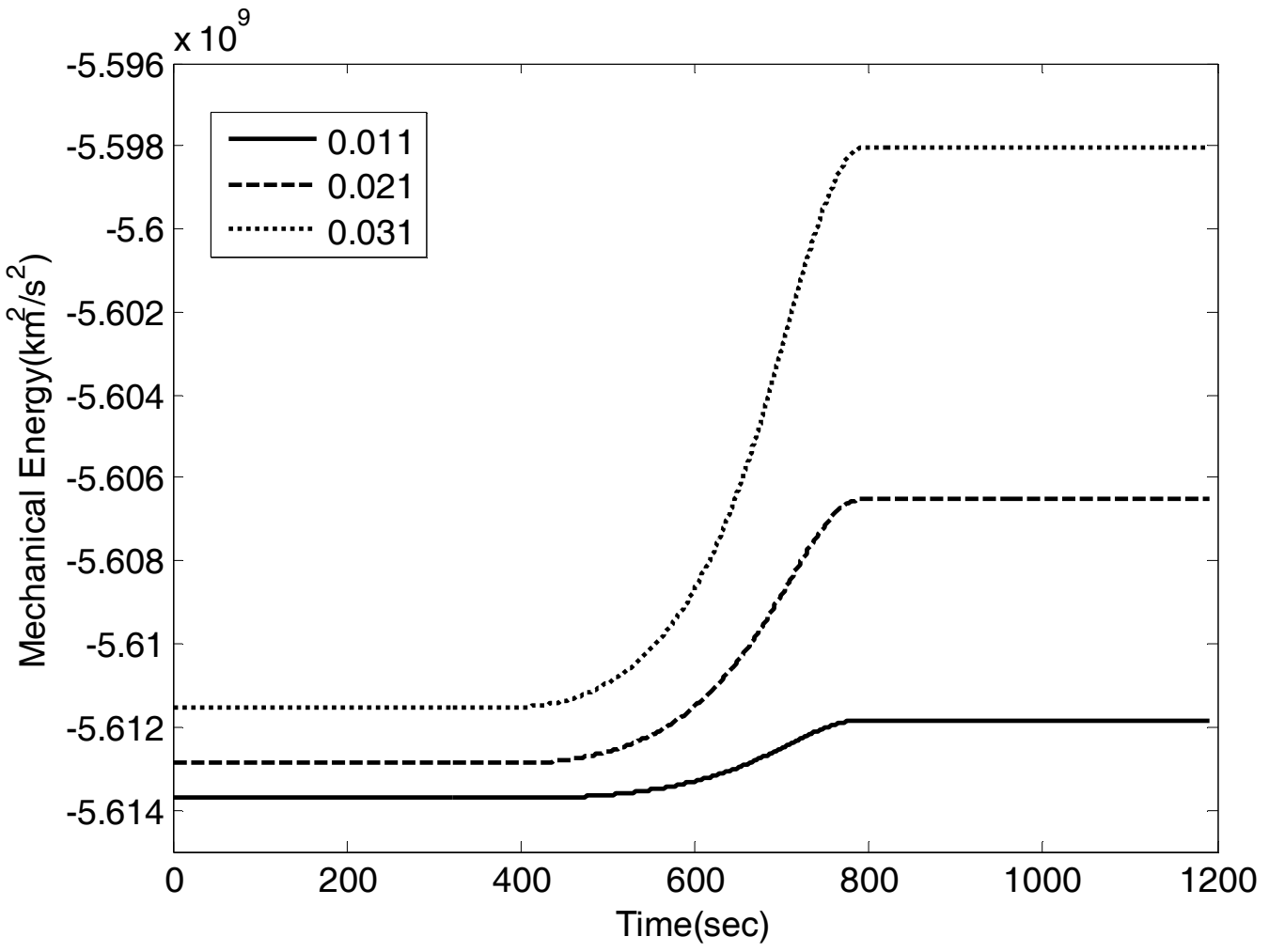

**Fig. 6 Simulation results of the local energy effects. The bottom graph shows the mechanical energy increments with different spinning angular rates: 0.011 rad/s, 0.021 rad/s, 0.031 rad/s.**

As Fig. 6 illustrated, calculating results shown the energy variation when the tether retrieves from 10 km to 4 km. Accordingly, the mechanical energy takes abrupt increase and more increment is achieved with greater spinning angular momentum.

## B. Angular Momentum Transportation

Since the total angular momentum of the system is conserved and the transportation between the spinning angular momentum and orbital angular momentum always exists, it is of great interest to inspect the dynamic mechanism of this phenomenon, especially for that with the tether length variable. The equations of motion are analyzed and there are several points that are helpful to make the lift hand side of equation (4) is the radial component of mass center acceleration, and its right hand side is the summation of gravitational forces' radial components; the lift hand side of equation (5) is the transverse component of mass center acceleration, and its right hand side is the summation of gravitational forces' transverse components; the lift hand side of equation (6) is the derivative of spinning angular momentum, and its right hand side is the summation of moment of gravitational forces relative to the mass center.

For a fast spinning tether that we concerned, i.e. $|\dot{\alpha}| \gg |\dot{\varphi}|$, the right hand side of (4) has invariant sign, which is always negative. However, the sign of right hand side of (5) (or of (6)) changes as $\rho$ , $\varphi$ , $\alpha$ and $l$ . Term $\left\{ \frac{1}{\left[\rho^2 - 2\rho\lambda_2 l\cos\alpha + (\lambda_2 l)^2\right]^{3/2}} - \frac{1}{\left[\rho^2 + 2\rho\lambda_1 l\cos\alpha + (\lambda_1 l)^2\right]^{3/2}} \right\}\sin\alpha$ causes sign change, meaning that when $\alpha = k\pi$ or $|OS_1| = |OS_2|$, the signs of right hand sides of (5) and (6) change. To inspect one circle, $\alpha = 0 \sim 2\pi$, there are 4 critical points: $0$, $\alpha_1$, $\pi$, $\alpha_2$, that the sign of right hand side of (5) changes as following.

**Table 1. Sign of right hand side of (5)**

| $0 \sim \alpha_1$ | $\alpha_1 \sim \pi$ | $\pi \sim \alpha_2$ | $\alpha_2 \sim 2\pi$ |
|---|---|---|---|
| + | - | + | - |

And the sign of right hand side of (6) is always opposite to that of (5). It is substantially coincident with the conservation of total angular momentum. The values of $\alpha_1$ and $\alpha_2$ are determined by

$$\cos\alpha = \frac{(\lambda_2 - \lambda_1)l}{2\rho} \tag{20}$$

Noticing $\rho \gg l > 0$, then approximately, we get $\alpha_1 \approx \pi/2$ and $\alpha_2 \approx 3\pi/2$.

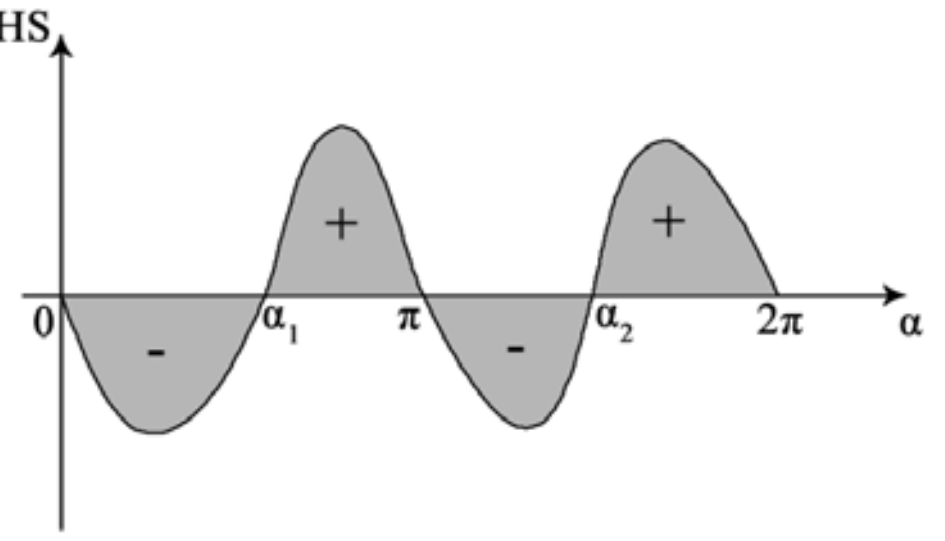

**Fig. 7 The right hand side of equation (6) varying as spin angle** $\alpha$

Fig. 7 demonstrates the variation of right hand side of equation (6) as spin angle $\alpha$, implying that the spinning angular momentum and orbital angular momentum alternate ceaselessly. Combining with the fact that $\alpha$ increases monotonely as time $t$, $\alpha$ could be seen as new regular parameter instead of $t$, which preserves the equivalent analysis by a bicontinuous mapping.

Defining

$$V = \frac{-\mu\rho\sin\alpha}{\left[\rho^2 - 2\rho\lambda_2 l\cos\alpha + (\lambda_2 l)^2\right]^{3/2}} + \frac{\mu\rho\sin\alpha}{\left[\rho^2 + 2\rho\lambda_1 l\cos\alpha + (\lambda_1 l)^2\right]^{3/2}} \tag{21}$$

The transferred angular momentum in one circle is determined by $\Delta L = \int_0^{2\pi} V(\rho,\alpha,l)d\alpha$, which is demonstrated with the summation of shadow areas in Fig. 7. Our interest is to dominate $\Delta L$ during one spinning period, and considering the widths of ranges $0 \sim \alpha_1$, $\alpha_1 \sim \pi$, $\pi \sim \alpha_2$, $\alpha_2 \sim 2\pi$ are almost equal, the approach is to regulate the extreme values of $V$.

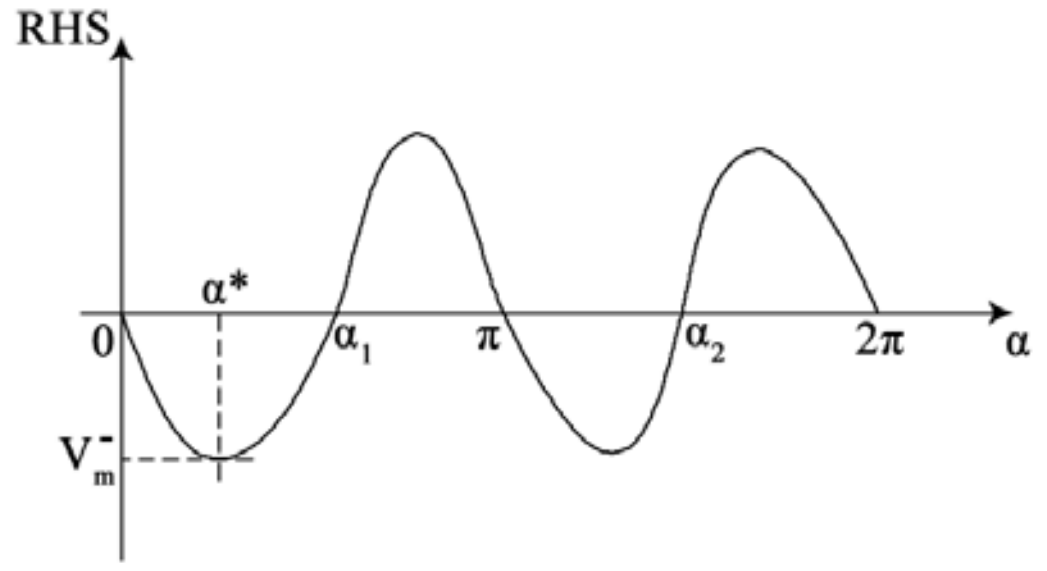

**Fig. 8 The extreme value of $V$ in the range $(0,\alpha_1)$**

As illustrated in Fig. 8, considering the extreme point in the range $(0,\alpha_1)$,

$$\frac{\partial}{\partial\alpha}V=\cos\alpha\left(\frac{-1}{|OS_1|^{3/2}}+\frac{1}{|OS_2|^{3/2}}\right)+3\rho l\sin^2\alpha\left(\frac{\lambda_2}{|OS_1|^{5/2}}+\frac{\lambda_1}{|OS_2|^{5/2}}\right)=0$$

Where $|OS_1|=\left[\rho^2-2\rho\lambda_2 l\cos\alpha+(\lambda_2 l)^2\right]^{1/2}$ and $|OS_2|=\left[\rho^2+2\rho\lambda_1 l\cos\alpha+(\lambda_1 l)^2\right]^{1/2}$. The extreme value

$$V_m^-=\mu\rho\sin\alpha^*\left(\frac{-1}{|OS_1|^{3/2}}+\frac{1}{|OS_2|^{3/2}}\right)=\frac{-3\mu\rho^2 l\sin^3\alpha^*}{\cos\alpha^*}\left(\frac{\lambda_2}{|OS_1|^{5/2}}+\frac{\lambda_1}{|OS_2|^{5/2}}\right)$$

Due to $\rho\gg l>0$, the term $\left(\frac{\lambda_2}{|OS_1|^{5/2}}+\frac{\lambda_1}{|OS_2|^{5/2}}\right)\approx\frac{1}{\rho^{5/2}}$, meaning that the value of $V_m^-\approx\frac{-3\mu l\sin^3\alpha^*}{\rho^{1/2}\cos\alpha^*}$ depends on the tether length $l$. In order to make the spinning angular momentum increase or decrease in one spinning period, the tether length reaction must be consistent with $\alpha$ quadrants, which is listed as follows.

**Table 2. Control law of tether reaction**

| $\Delta L$ | $0\sim\alpha_1$ | $\alpha_1\sim\pi$ | $\pi\sim\alpha_2$ | $\alpha_2\sim 2\pi$ |
|---|---|---|---|---|
| + | 1+ | 1- | 1+ | 1- |
| - | 1- | 1+ | 1- | 1+ |

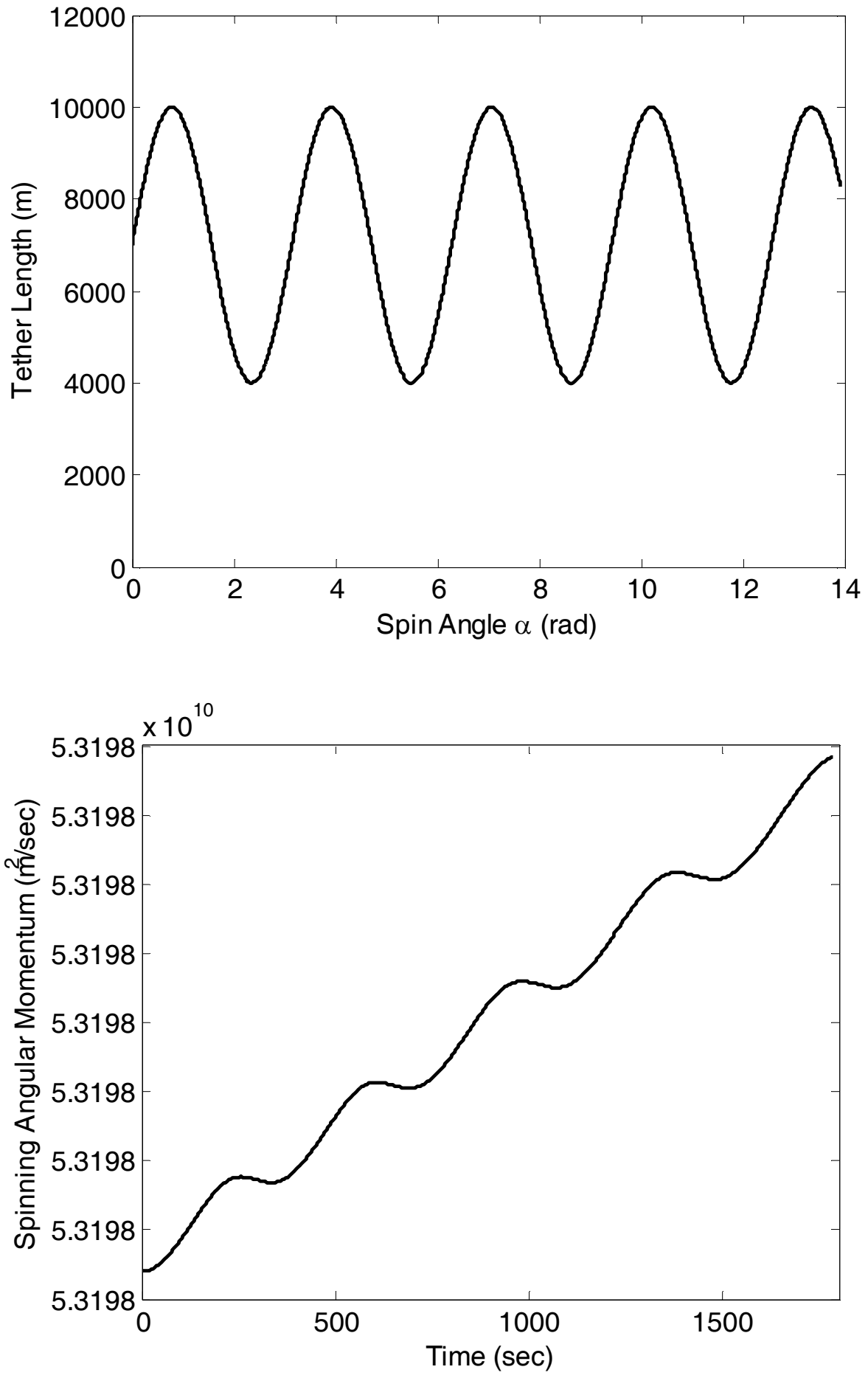

**Fig. 9 Simulation results of the tether reaction to increase the spinning angular momentum, with the tether length changes between 4 km and 10 km**

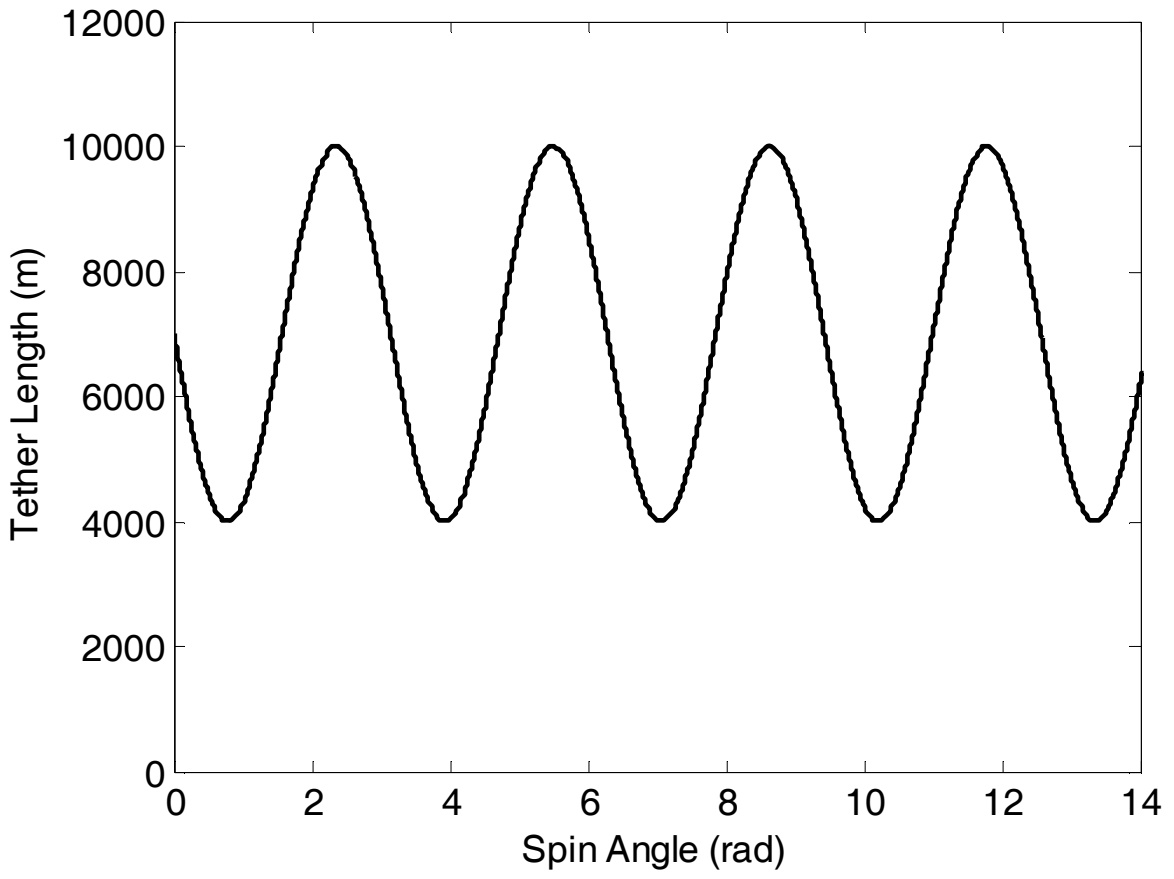

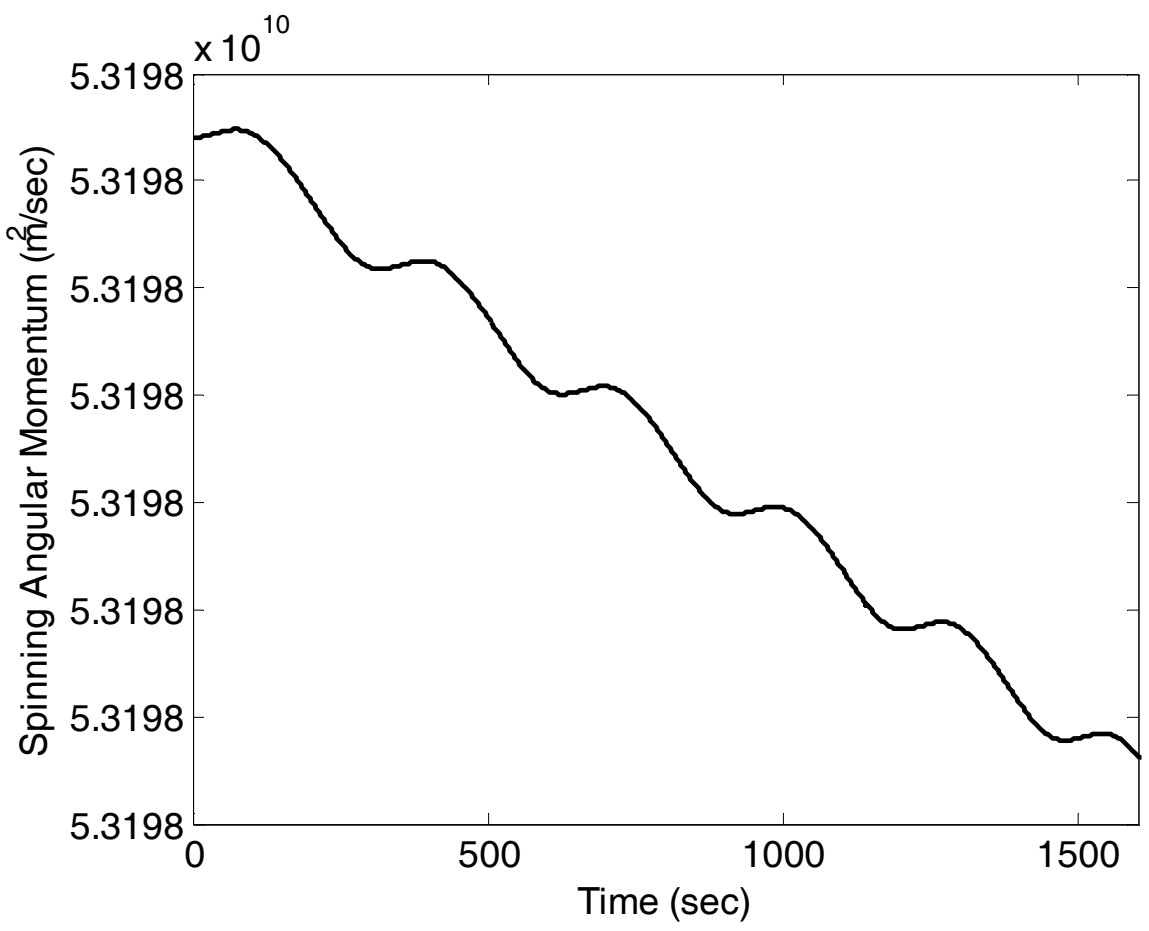

**Fig. 10 Simulation results of the tether reaction to decrease the spinning angular momentum, with the tether length changes between 4 km and 10 km**

Fig. 9 illustrated the spinning angular momentum increase when the tether deploys at $(0, \alpha_1)$, $(\pi, \alpha_2)$ and retrieves at $(\alpha_1, \pi)$, $(\alpha_2, 2\pi)$. Accordingly, Fig. 10 illustrated the spinning angular momentum increase when the tether deploys at $(\alpha_1, \pi)$, $(\alpha_2, 2\pi)$ and retrieves at $(0, \alpha_1)$, $(\pi, \alpha_2)$. The simulation proves that the angular momentum transferred between orbit and spinning is under control with this method, and combining the relationship between the mechanical energy increments and spinning rate, the local energy increment could also be regulated with this approach employed.

## V. Orbital Propulsion via Tether Reaction

Since specific discussions are promoted on local effects of tether reaction and long-term orbital motion in previous sections, it is supported that orbital propulsion could be achieved via tether reaction. This section deals with the control laws of orbital motion through deploying and retrieving the tether. Because the orbit maneuvres in this way is extremely slow, the mean variation of orbit shape here is investigated with a perturbation method employed[11].

The resultant force $F_C$ depends on tether length $l$ and polar radius $\rho$. The value of $\rho$ oscillates at orbital period, while tether length $l$ may change much faster than $\rho$ that the reaction time could be ignored compared with the orbital period. With consideration of $\rho \gg l$, the calculation results show that the value of $F_C$ is slightly disturbed from the two-body gravitational force $-\mu/\rho^2$, therefore the perturbation model was introduced to the discussion of orbital propulsion. Define the perturbation force

$$F_r = F_C + \frac{\mu}{\rho^2} \tag{22}$$

The Lagrange's planetary equations are given in orbital elements, considering the perturbation force is radial and orbital motion is delimited in the plan, the reduced perturbation equations are

$$\dot{a} = \frac{2e\sin f}{n\sqrt{1-e^2}} F_r \tag{23}$$

$$\dot{e} = \frac{\sqrt{1-e^2}\sin f}{na} F_r \tag{24}$$

$$\dot{\omega} = -\frac{\sqrt{1-e^2}\cos f}{nae} F_r \tag{25}$$

$$\dot{M} = n - \frac{1-e^2}{nae}\left(\frac{2e\rho}{p} - \cos f\right) F_r \tag{26}$$

### A. Orbital Propulsion

Since the effects of tether reaction on the orbital motion is weak and slow, it takes time to produce considerable orbit change. The perturbation equations (23)-(26) are averaged in order to measure the slight mean difference of orbit shape during a long term. For the orbital element $\sigma$, equations (23)-(26) are integrated as $\dot{\sigma}_{avg} = \frac{1}{T}\int_0^T \dot{\sigma} dt$, and with $dt = \sqrt{p^3} \cdot \frac{df}{(1+e\cos f)^2}$ substituted, the mean perturbation equations are

$$\dot{a}_{avg} = \frac{e\left(1-e^2\right)}{\pi n}\int_0^{2\pi} \frac{\sin f \cdot F_r}{\left(1+e\cos f\right)^2} df \tag{27}$$

$$\dot{e}_{avg} = \frac{\left(1-e^2\right)^2}{2\pi na}\int_0^{2\pi} \frac{\sin f \cdot F_r}{\left(1+e\cos f\right)^2} df \tag{28}$$

$$\dot{\omega}_{avg} = \frac{\left(1-e^2\right)^2}{2\pi nae}\int_0^{2\pi} \frac{-\cos f \cdot F_r}{\left(1+e\cos f\right)^2} df \tag{29}$$

It is clear that $\dot{a}_{avg}/\dot{e}_{avg} = 2ea/\left(1-e^2\right)$, thus the variation of semi-major axis and eccentricity are always synchronized. Comparing the right hand side of (27) with that of (29), the integral functions are orthogonal. It suggests that the semi-major axis and eccentricity always vary consistently and the argument of perigee can be controlled separately.

Since the mean orbit shape is determined by $a_{avg}$, $e_{avg}$ and $\omega_{avg}$, the efficiency of propulsion using tether reaction is estimated with equations (27)-(28). By (22), $F_r$ is positively correlated to tether length, so the tether length boundaries should be reached to maximize the propulsion capacity in some aspect. Specifically, the extreme values of $\dot{a}_{avg}$, or $\dot{e}_{avg}$, are reached when tether length varies as follow.

**Table 3. Control law of tether reaction**

| Case | Reaction location | Tether length |
|---|---|---|
| $\left[\dot{a}_{avg}\right]_{\max}$ | $f$ =0 | $l_{max}$→$l_{min}$ |
| | $f$ =π | $l_{min}$→$l_{max}$ |
| $\left[\dot{a}_{avg}\right]_{\min}$ | $f$ =0 | $l_{min}$→$l_{max}$ |
| | $f$ =π | $l_{max}$→$l_{min}$ |

Fig. 11-13 illustrated the variation of orbit elements with this control law, in which the model parameters given beforehand. The tether length is between 20 km and 180 km, and initial spin rate is 0.051 rad/s. As Fig. 11 shown, the semi-major axis increases acceleratedly, while Fig. 12

displayed an approximately linear increase of the eccentricity ratio. Fig. 13 showed the orbit drifts clockwise at slow rate.

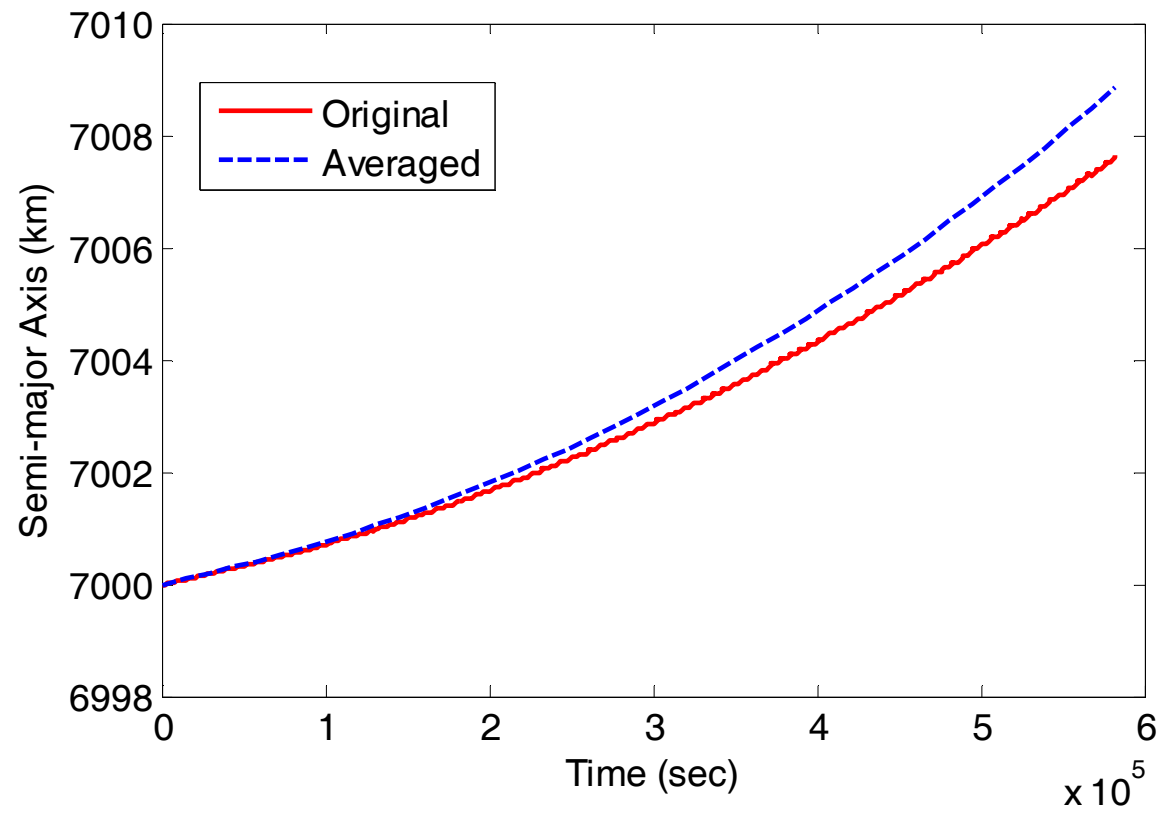

**Fig. 11 The simulation of semi-major axis varying as time, based on original model and averaged model. Since the increasing is accelerated, at the beginning, average rate of the semi-major axis is about 50 meters per hour**

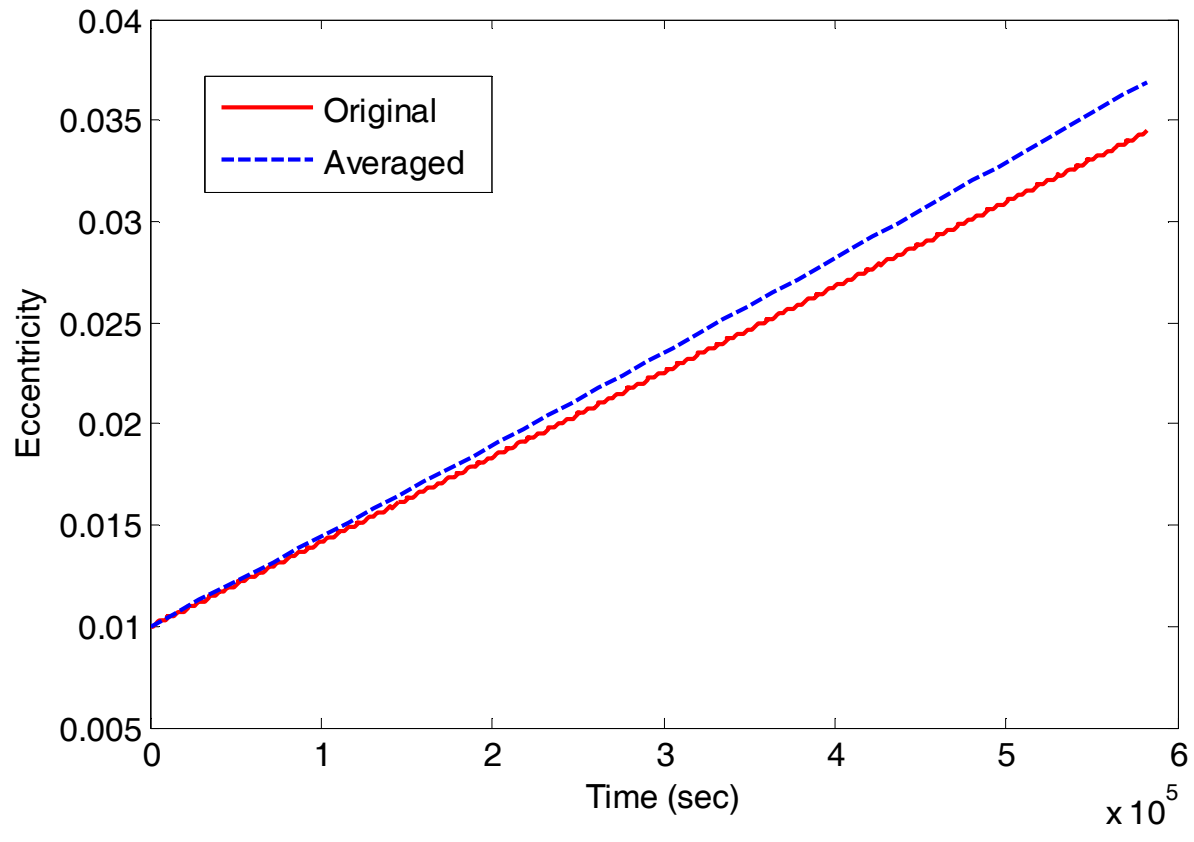

**Fig. 12 The simulation of eccentricity varying as time, based on original model and averaged model. The relationship of the eccentricity ratio over time is approximately linearity**

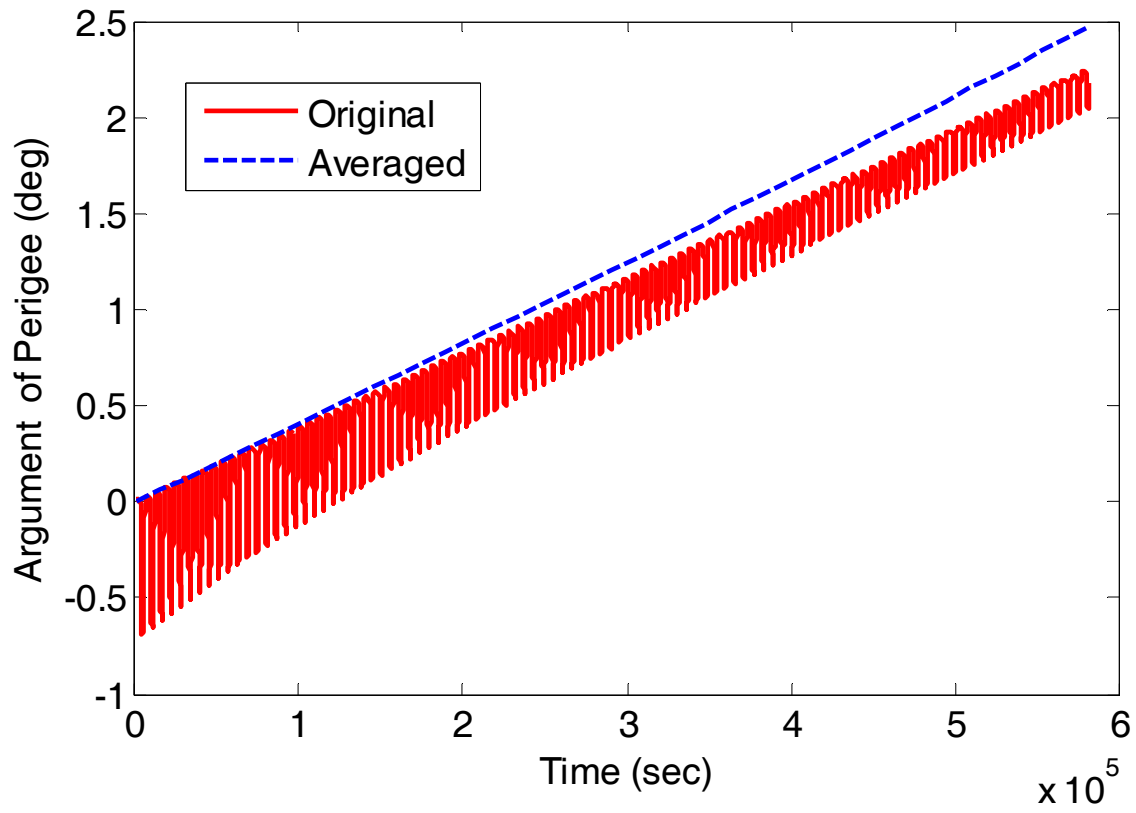

**Fig. 13 The simulation of the argument of perigee varying as time, based on original model and averaged model. The drift rate is quite small and approximately constant as 4.0×10$^{-6}$ degree/s.**

Likewise, the extreme values of $\dot{\omega}_{avg}$ are achieved when tether length varies as follow.

**Table 4. Control law of tether reaction**

| Case | Reaction location | Tether length |
|---|---|---|
| $\left[\dot{\omega}_{avg}\right]_{max}$ | $f=\pi/2$ | $l_{max}\rightarrow l_{min}$ |
| | $f=3\pi/2$ | $l_{min}\rightarrow l_{max}$ |
| $\left[\dot{\omega}_{avg}\right]_{min}$ | $f=\pi/2$ | $l_{min}\rightarrow l_{max}$ |
| | $f=3\pi/2$ | $l_{max}\rightarrow l_{min}$ |

Fig. 14-16 illustrated the variation of orbit elements with this control law, in which the model parameters given beforehand. As above discussed, the semi-major axis and eccentricity ratio keep invariant averagely, and the results on argument of perigee consistently indicate linear increase with time.

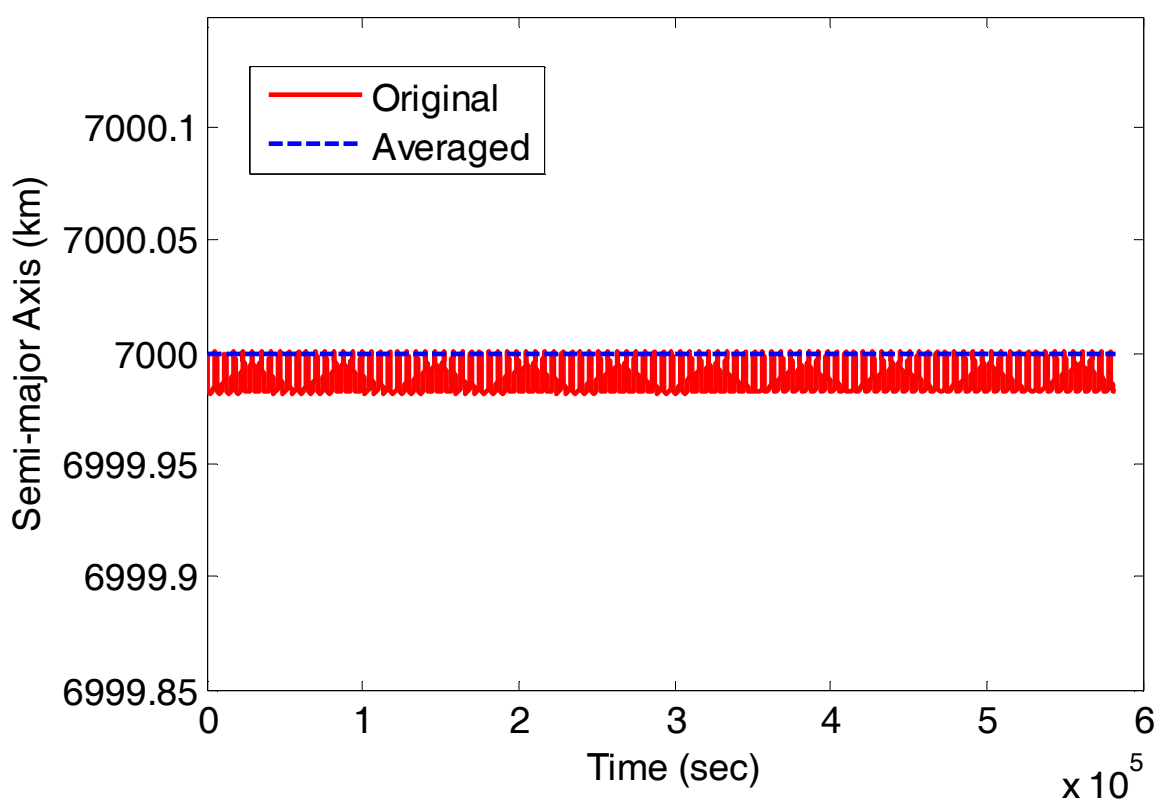

Fig. 14 The simulation of the semi-major axis varying as time, based on original model and averaged model. The consistent results show the mean semi-major axis keeps constant

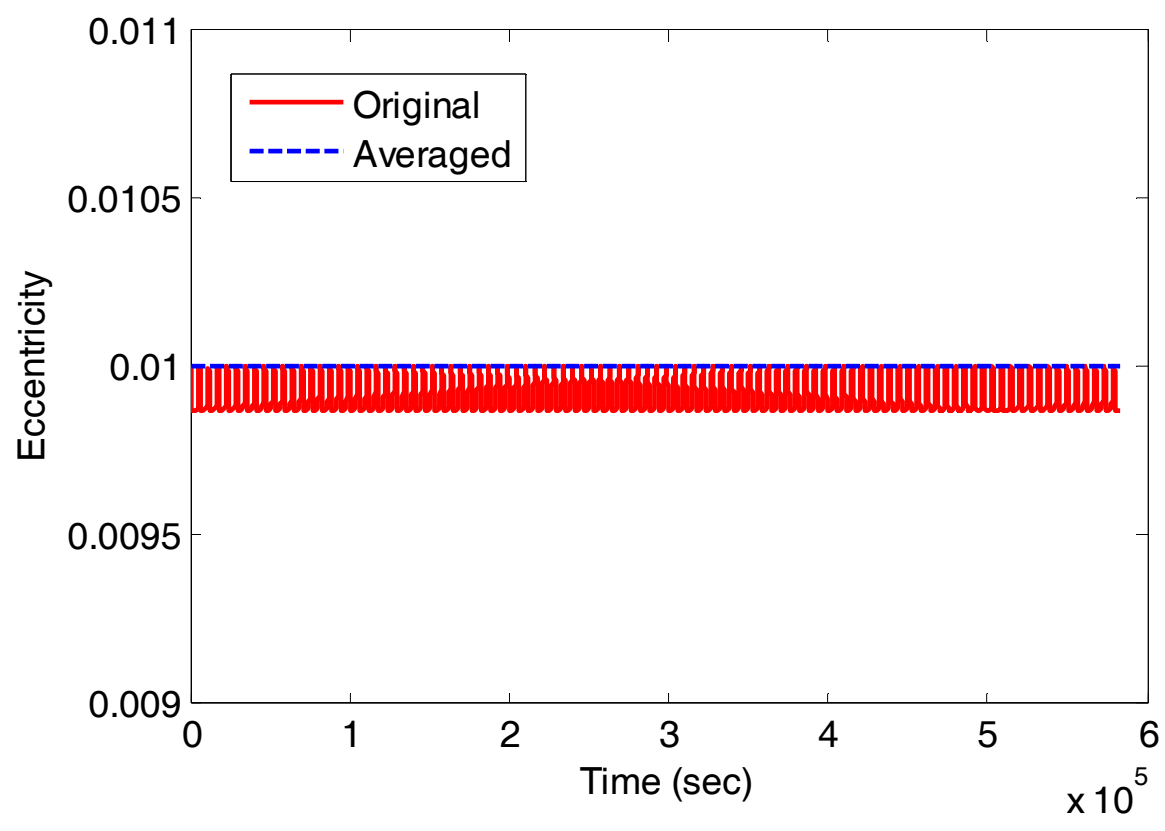

**Fig. 15 The simulation of the eccentricity varying as time, based on original model and averaged model. The consistent results show the mean eccentricity keeps constant**

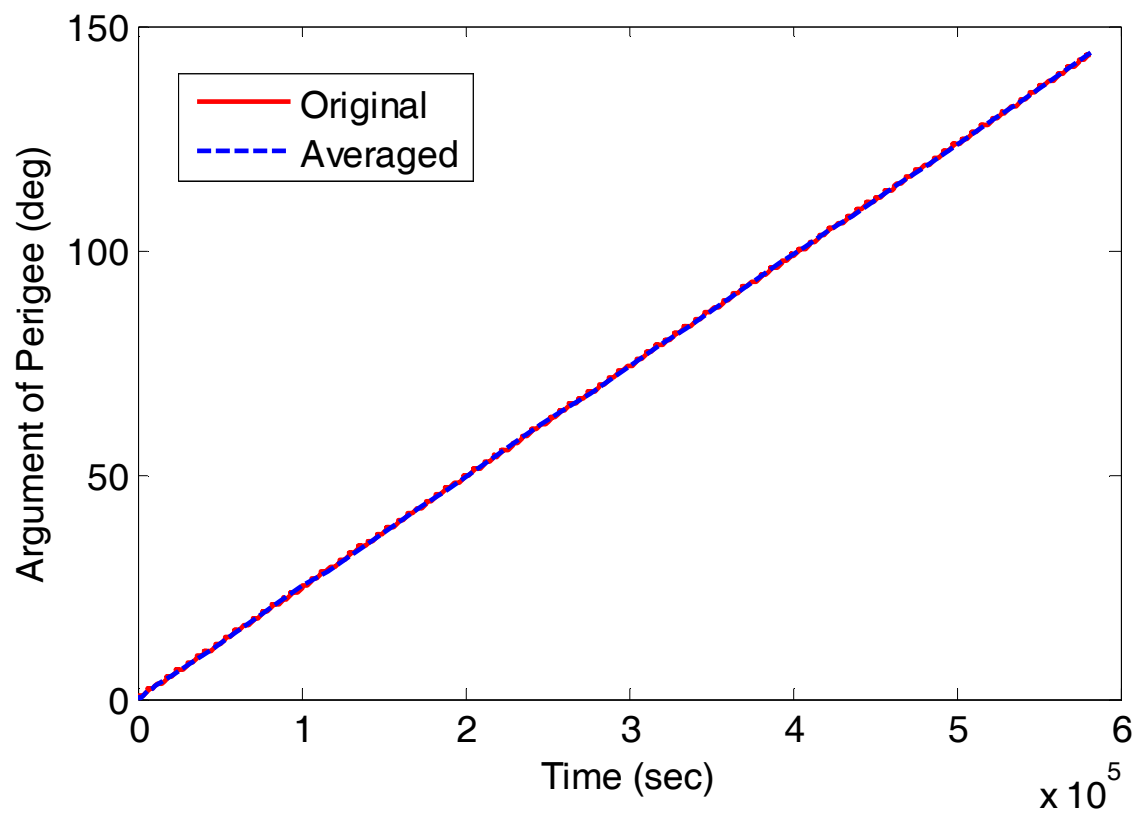

**Fig. 16 The simulation of the argument of perigee varying as time, based on original model and averaged model. The consistent results show a linear increase of the argument of perigee, i.e. the orbit drifts at a uniform angular speed**

The results suggest that the efficiency of propulsion for the planar orbits of mass center, viewing from long term, the tether reaction of spinning tethered binary system could be used for propulsion. Additionally, the semi-major axis and eccentricity are synchronized and the argument of perigee could be controlled independently.

## B. Orbit Maintaining

As Fig. 17 illustrated, specific operation of the tether length could be applied to freezing the orbit shape. In particular, maintaining constant perturbation acceleration $F_r$ leads to invariant mean semi-major axis and eccentricity ratio and clockwise drifting argument of perigee.

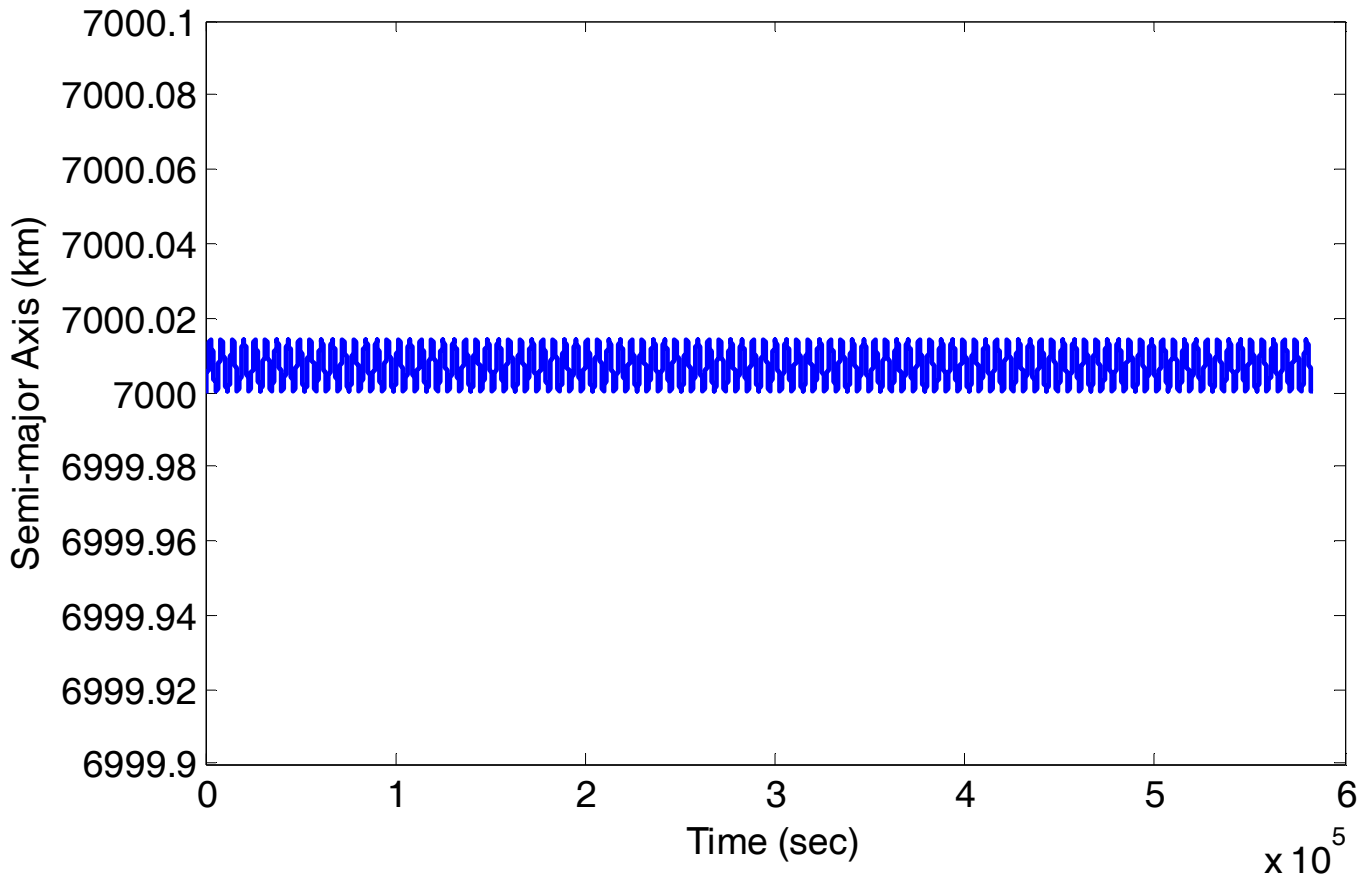

**Fig. 17 The simulation of the semi-major axis varying as time, based on original model and with the constant perturbation acceleration**

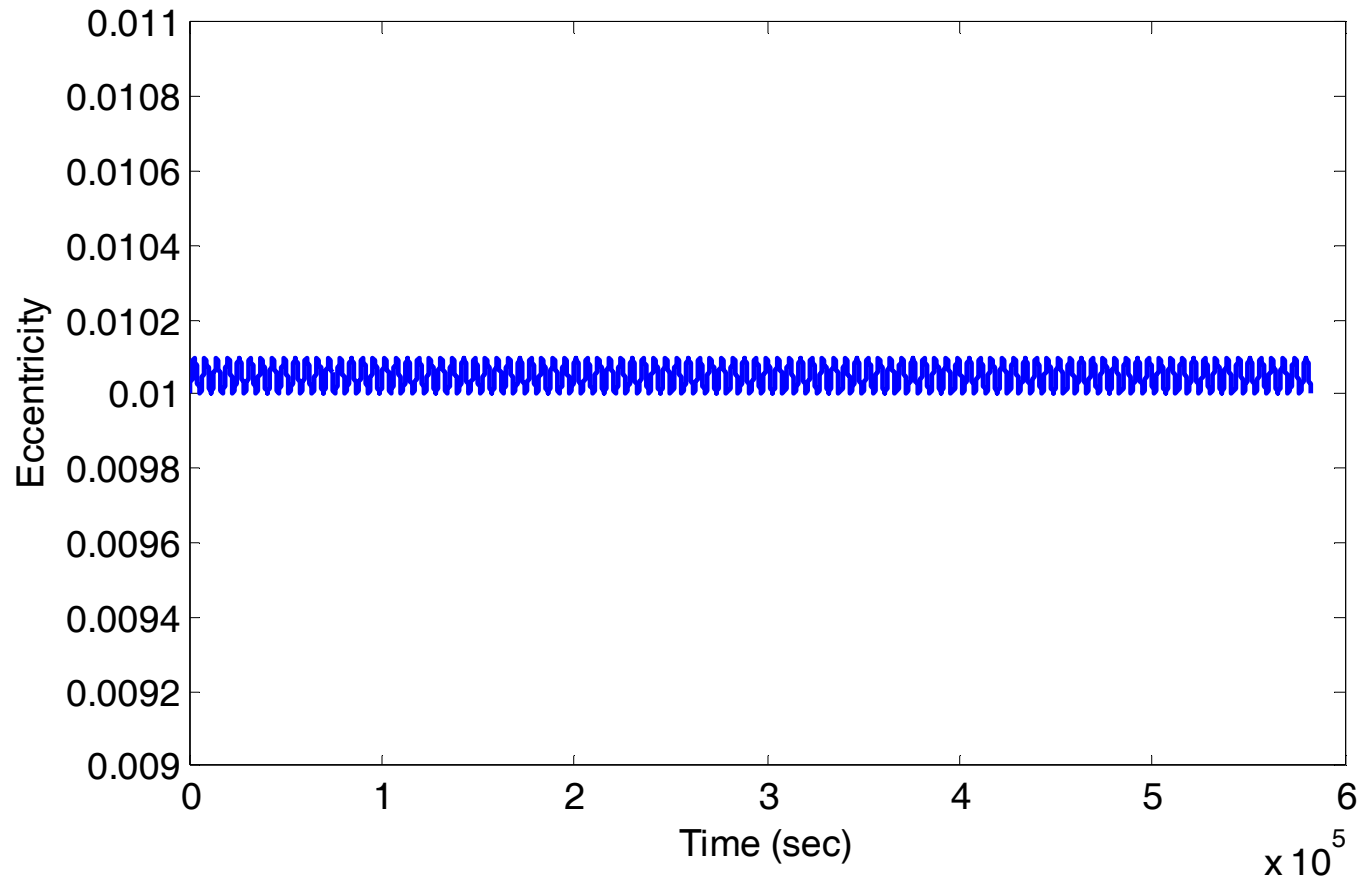

**Fig. 18 The simulation of the eccentricity varying as time, based on original model and with the constant perturbation acceleration**

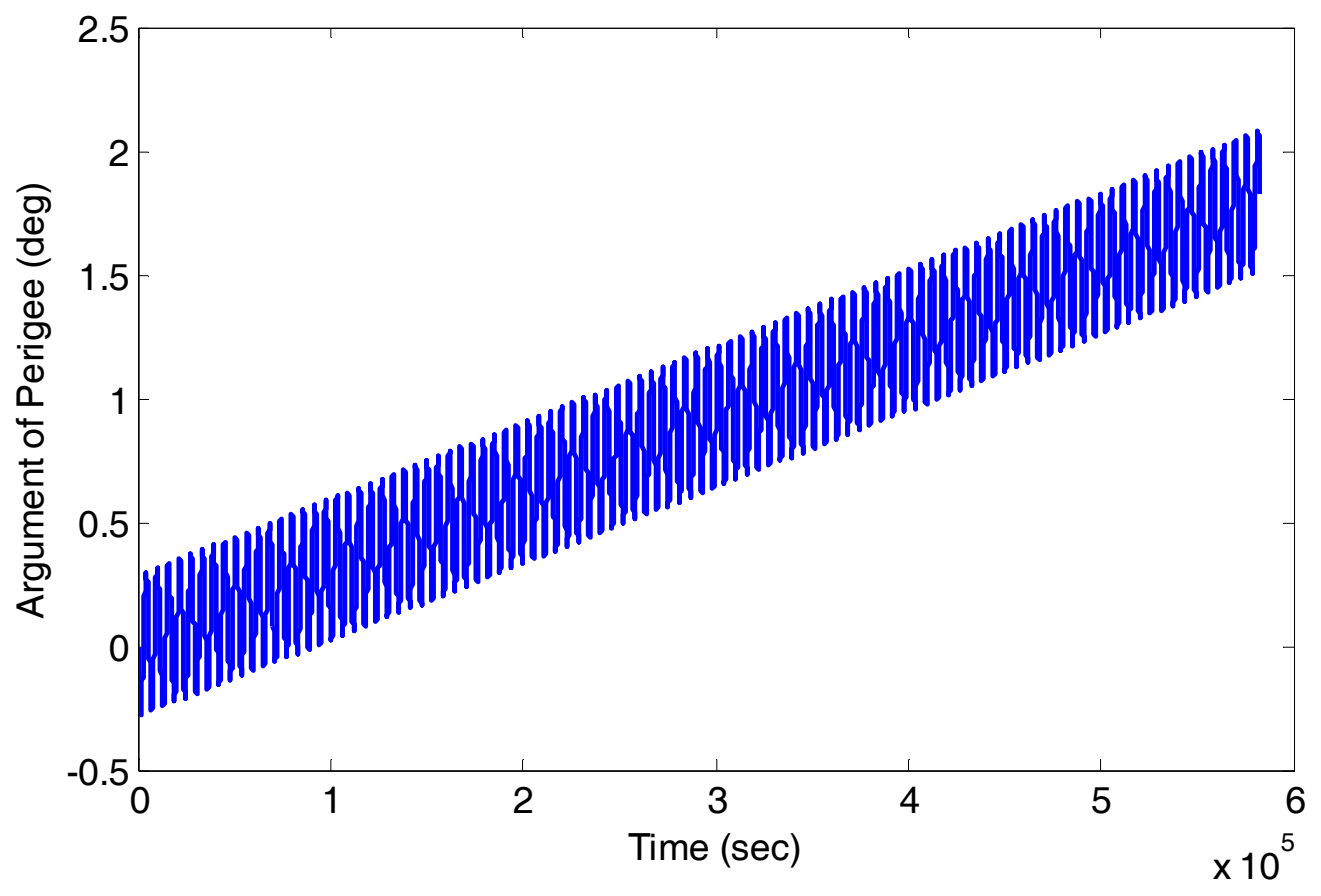

**Fig. 19 The simulation of the argument of perigee varying as time, based on original model and with the constant perturbation acceleration**

Likewise, a mean frozen orbit could be obtained, i.e. the mean orbital elements $a_{avg}$ , $e_{avg}$ and $\omega_{avg}$ keep stationary. The approach is to set the perturbation acceleration $F_r$ subjecting to $F_r = A(1+e\cos f)^2$, where $A$ is const satisfying $-F_{\max} < F_r < -F_{\min}$. Fig. 20-22 illustrates the orbit elements varying as time.

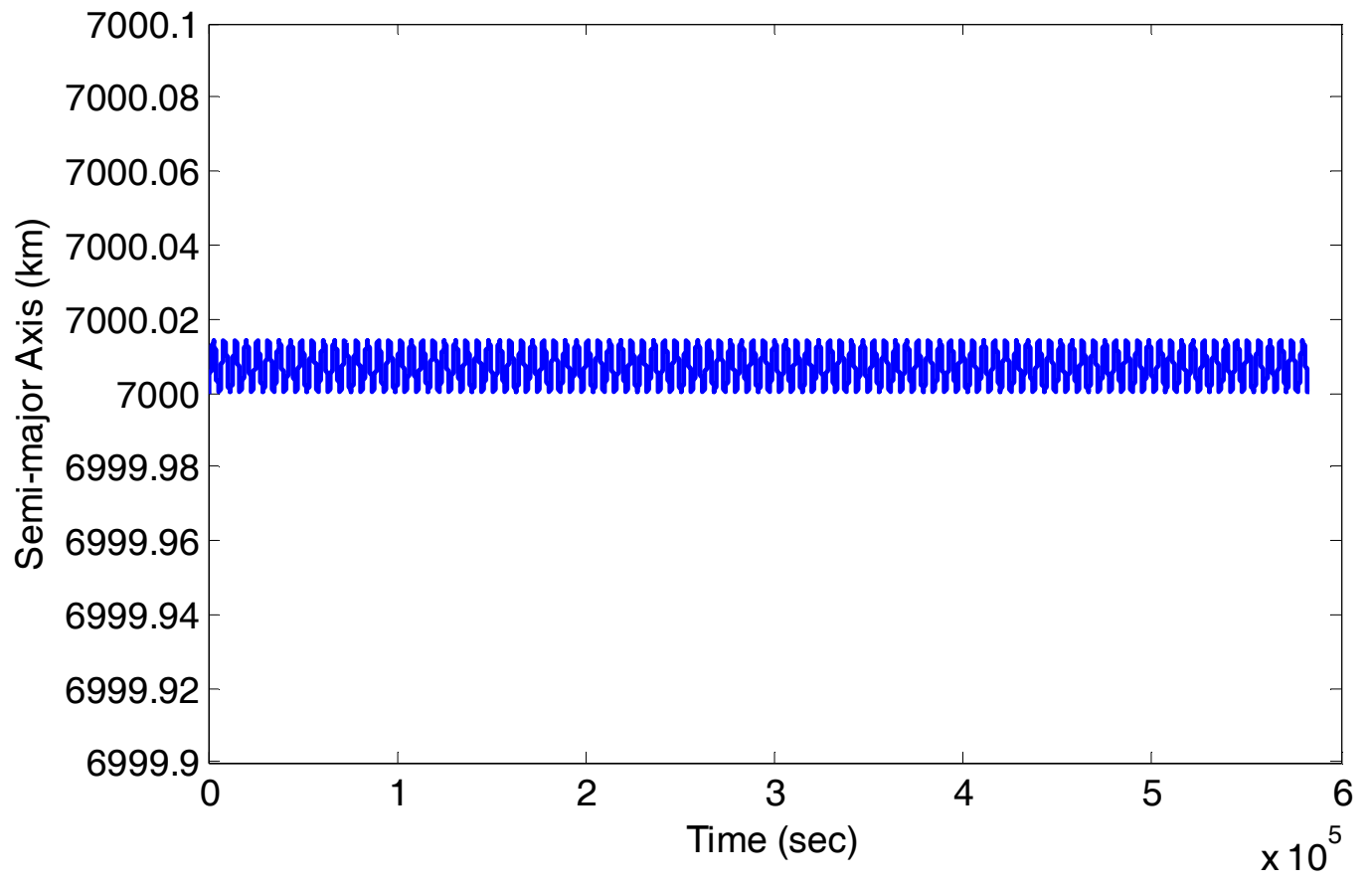

**Fig. 20 The simulation of the semi-major axis varying as time, based on original model and with above control laws**

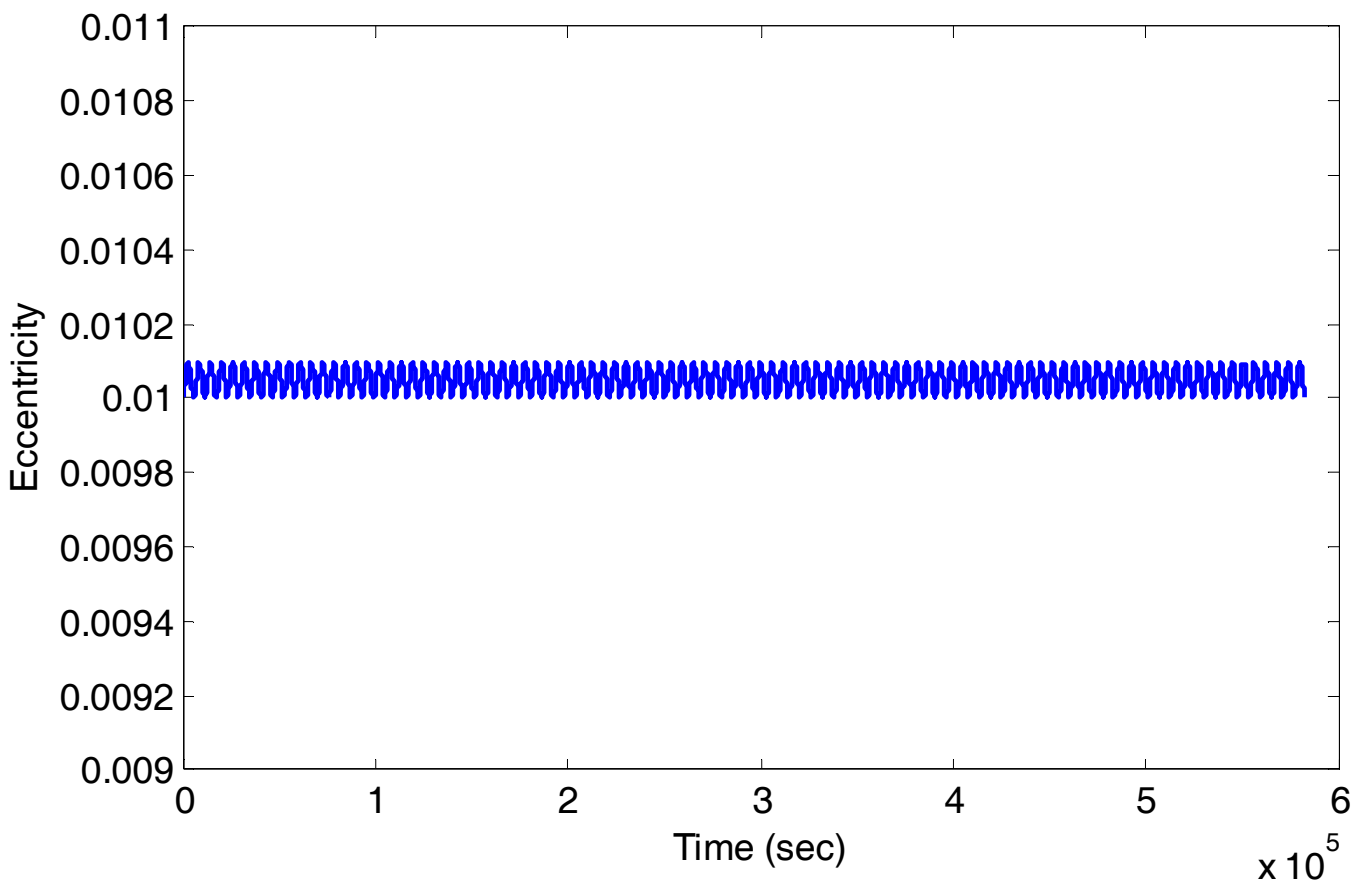

**Fig. 21 The simulation of the eccentricity varying as time, based on original model and with above control laws**

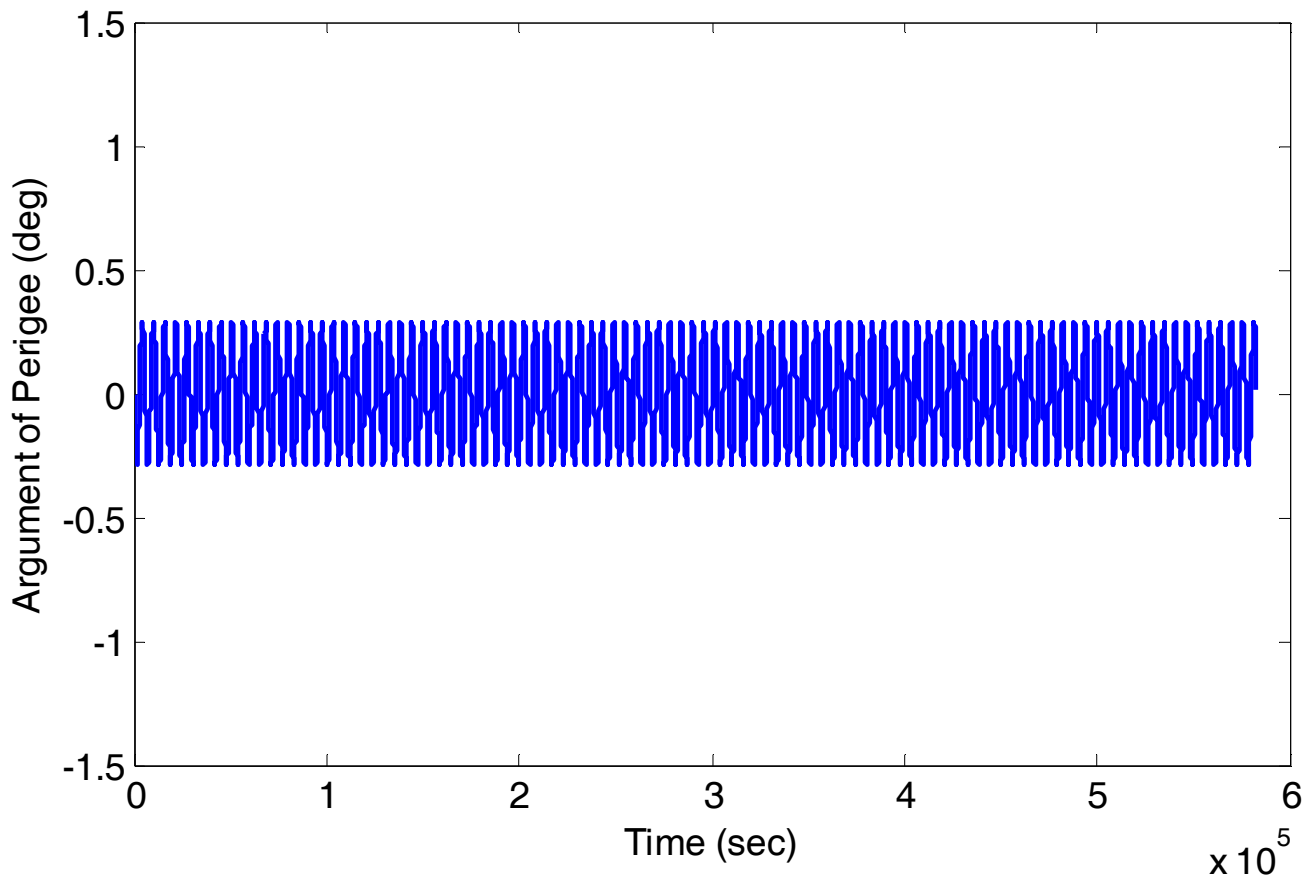

**Fig. 22 The simulation of the argument of perigee varying as time, based on original model and with above control laws**

## VI. Conclusions

The present work focuses on the phenomenon that orbit propulsion could be simply resulted from regular inertia changes. The spinning tethered binary system is employed to analyze the mechanism of this phenomenon. The general orbital motion of system mass center is determined analytically by separating the motion of slow variants from original system, which is a quasi-periodic orbit with the perigee drifting uniformly. The local study on the energy effects of tether reaction shows that high reaction attitude and high spin rate generate great energy increment. Besides, we studied the angular momentum transportation between the orbital motion and spinning motion and developed an approach of changing tether length to implement this transportation effectively. Averaged perturbation method is applied to reveal the secular effect of orbital maneuver, which is numerically verified for specific tasks later. Several conclusions could be obtained: this propulsion method is relatively slow, and the propulsion efficiency depends on the tether strength and the power of the winch on satellite; the semi-major axis and eccentricity of mass center orbit, which are always synchronized, could be controlled simultaneously, and the argument of perigee is independent of them. Finally, some applications of this propulsion method are simulated and discussed, such as freezing part or all of the orbit elements to maintain the orbit shape.